# Decentralized Knowledge and Learning in Strategic Multi-user Communication

Yi Su and Mihaela van der Schaar

Dept. of Electrical Engineering, UCLA

{yisu, mihaela}@ee.ucla.edu

## I. INTRODUCTION

*A. Motivation*

Multi-user wireless communications systems form *competitive* environments, where *heterogeneous* and *strategic* users compete for the available spectrum resources. These heterogeneous devices may differ in terms of their adopted standards and architectures, their deployed communication algorithms, their experienced environment (e.g. channel conditions and traffic characteristics), their application-defined utilities etc. [1]. Moreover, these devices can also significantly differ in their ability to sense the environment and acquire information about other users sharing the same spectrum, exchange information and negotiate spectrum access rules with the competing users and, ultimately, learn and reason based on the available information to select their optimal transmission strategies. Importantly, in most communication scenarios, users compete for resources in a strategic manner, i.e. they aim at maximizing their own utilities.

The majority of past multi-user communications research has focused on analyzing and quantifying the performance of various multi-user environments, where transceivers with similar standards, algorithms or utilities share the available spectrum resources. An underlying assumption has been that the users select their transmission actions based on *either complete or no knowledge about their competitors*' protocols, strategies, utility functions, environment etc. In this report, the term "knowledge" represents the correct beliefs [1] of users about their opponents and the environment.

---

[1] Plato defined knowledge as "justified true belief" [http://en.wikipedia.org/wiki/knowledge].

Specifically, two extreme types of multi-user interaction have been especially studied in the past:

- the complete knowledge scenario, which assumes that all participating users know[2] the other users' actions, utility forms etc.;

- the private knowledge (e.g. the knowledge of its own channel or traffic characteristics) scenario, where users interact by assuming no knowledge about each other.

Hence, the aforementioned communications scenarios neglect that in a multi-user communication system, the users' knowledge is mostly decentralized and heterogeneous.

Another assumption in the existing research has been that there is consensus about what is a fair distribution of resources and thus, efficiency and fairness can be simultaneously satisfied. However, this violates the strategic nature of users and, as a result, it encourages a passive participation of transceivers in the multi-user interaction, because wireless users are not able to *proactively influence the resource division based on their knowledge*. Thus, such current multi-user network designs do not consider the heterogeneity and decentralization of the knowledge of the participating devices and, do not take advantage of the transceivers' "smartness" (i.e. their ability to acquire information, learn and reason about their opponents), which often lead in practice to inefficient spectrum usage. For instance, in existing centralized resource management solutions for 802.11e Hybrid Coordination Function, users operate in a non-collaborative manner and request spectrum access from the moderator (e.g. the access point or base station) by declaring their worst case resource needs, while disregarding the resource requests of their competing users [48].

Alternatively, in this report, we present relatively recent research developments aimed at building a new paradigm for multi-user communication environments, where the emerging interaction among users and their resulting performance will be driven by the heterogeneous users' ability to strategically adapt their transmission actions, based on their knowledge about their competitors and the environment. Hence, voluntary rather than externally imposed (e.g. protocol enforced) collaborations may evolve among strategic users, because such collaborations may allow them to increase their own utility [44]. The development of such new paradigms for multi-user communications scenarios is essential, given the recent proliferation of heterogeneous protocols, devices, applications and services, but also diverse user needs.

[2] The assumption is that this knowledge was obtained by users publicly announcing their private information, or by implicitly assuming that all users that are co-existing adopt the same protocol [7][13].

*B. A user-centric approach for multi-user communications*

In this report, we will not adopt the conventional view of characterizing the multi-user interactions from a system designer's perspective, which aims to maximize the social welfare of the users by assuming that there is common knowledge about the users' preferences etc. and that there is consensus about how to define the welfare [45][46]. Instead, we will take the viewpoint of self-interested users, which have their own preferences and knowledge, and present a theoretical foundation that enables devices to proactively improve their own utilities by acquiring knowledge based on their strategic interaction with competing users. Note that this user-centric approach can also positively impact the overall communication system performance, as users are now proactively interacting and improving their knowledge. This decentralized knowledge paradigm is well-suited in the considered heterogeneous and distributed communication environments, where there is little or no available common knowledge.

In the presented paradigm for multi-user wireless systems, the heterogeneous transceivers are modeled as self-interested agents that strategically interact by adapting their transmission actions in order to maximize their own payoffs. The transmission actions are the algorithms and configurations available in the protocols of various OSI layers (from the application to the physical layer). The heterogeneous users may differ not only in their available transmission actions, but also in their strategies for selecting their actions based on their utilities, and their available knowledge about other users and the environment., The emerging interaction among users and their resulting performance in the paradigm is driven not only by the users' ability to efficiently adapt their transmission strategies (e.g. their available modulation and coding schemes, ability to shape their traffic), but also by their ability to proactively acquire information and form accurate beliefs about their competitors and the environment. The knowledge that the users can build depends on their ability to acquire information (which may be restricted by their hardware constraints, adopted protocols, etc.) and their capability to reason and learn based on this information, in order to form beliefs about the other users' actions, policies, protocols etc. Using their knowledge, the users can predict the responses of their opponents to their actions, and hence optimize their strategies for selecting a specific action given their utility. Thus, the multi-user wireless interaction becomes a game played by strategic agents, based on their available actions, strategies and, additionally, their knowledge. In short, unlike in traditional communication theory, where the performance of each user and the system is influenced by the users' actions, protocols and experienced environment, in the communication paradigm discussed in this report, the knowledge based on which users select their actions will also play an essential role.

In this report, we show how different forms of knowledge about the environment and competing users can influence the actions selected by various users and, consequentially, shape the resulting multi-user interactions and the utilities obtained by the users. We also introduce strategic learning techniques that enable users to accumulate knowledge and maximize their utilities by acquiring information via their interactions with other users and, based on this information, building their beliefs about their competing users and the environment, and optimizing their transmission actions.

*C. Quantifying performance in multi-user environments with decentralized knowledge*

To characterize multi-user environments with decentralized knowledge, we discuss in this report two related metrics:

(a) *The value of knowledge.* First, we discuss which performance bounds can be attained by individual users, when users and/or resource moderators with different degrees of knowledge about the entire communication system and the competing users (e.g. the knowledge of other users' action space, strategies, utility functions, used protocols etc.) interact. We term these resulting performance bounds as "the value of knowledge". We show in this report how the heterogeneous users can select their actions, given their available knowledge.

(b) *The value of learning.* We also discuss methods for constructing operational solutions and algorithms, which enable each user to approach the performance bound. For this, users can systematically acquire information about other users and, based on this information, deploy strategic learning solutions that enable them to forecast the other users' responses and, ultimately, to optimize their transmission actions. We highlight several strategic learning techniques, which require various information overheads and complexity costs, and lead to different performance gains for the wireless users. To quantify these gains, we use a new metric, which we refer to as "the value of learning".

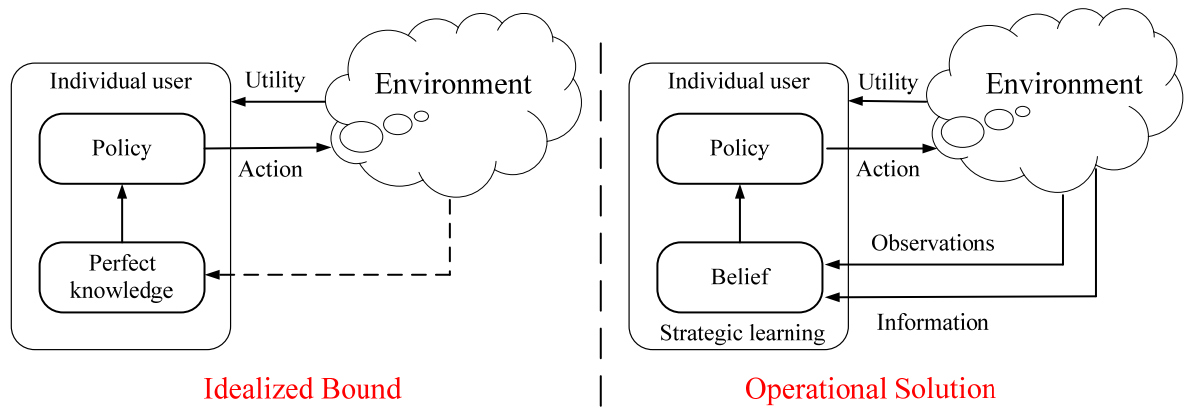

Fig. 1. Components of heterogeneous multi-user communication framework with decentralized knowledge.

As shown in Figure 1, while the value of learning depends on the practical capability of the user to reason and dynamically improve its beliefs based on its observations and explicit information exchanges, the value of knowledge represents the upper bound which can be achieved by a user, given its *observations* and *information*. The term observation refers to the observable outcome that users can access in the multi-user interaction without explicitly exchanging messages, while information is obtained by users through explicit message exchanges

Summarizing, the goal of this report is to provide the reader with a thorough understanding of how the users' *learning abilities*, their *knowledge*, and their *strategies* for selecting actions will impact and shape the multi-user *interaction* and the users' resulting *utilities*.

The report is divided into two main parts. First, we will discuss how the users can optimize their actions based on their available knowledge and show how to compute performance bounds achievable by the heterogeneous users with different knowledge. Second, to approach these bounds, we discuss how to construct operational algorithms for wireless users to proactively drive the information acquisition, and use strategic learning to form beliefs, based on which they can optimize their transmission strategies.

Section II describes the game theoretic approach in the communication setting and introduces a general mathematic model for communication games. Section III defines the value of available knowledge in communication games, formulates the optimal decision making for the users with heterogeneity in their available knowledge, and validates the performance improvement in power control games. In Section IV, the theory of learning in games is introduced as a tool for acquiring knowledge by users and improving their own performance. For instance, it is shown that in power control games, users can improve their performance by applying the regret matching algorithm. Section V concludes the report with the key challenges for future research.

## II. A Game Theoretic Model for Communication Networks

Game theory provides a formal framework for studying the interactions of strategic agents. Recently, there has been a surge in research activities that employ game theory to model and analyze a wide range of application scenarios in modern communication networks [2]-[4], including resource allocation [5][6], power control and spectrum management [7]-[13], medium access control [14]-[17], routing and congestion control [18]-[22], information theoretical analysis [23]-[25], etc. Depending on the characteristics of different applications, numerous game-theoretical models and solution concepts have been proposed to describe the multi-user interactions and optimize the users' decisions in communication networks.

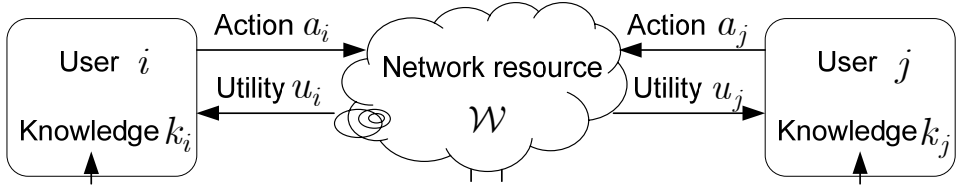

Fig. 2. A general illustration of knowledge-driven communication games

In this report, we first formulate the multi-user interaction in communication systems as a *knowledge-driven strategic game*, where users with different levels of knowledge availability compete for limited resources. As shown in Figure 2, the game is formally defined as a tuple $\Omega = Game\langle \mathcal{N}, \mathcal{W}, \mathcal{A}, \mathcal{U}, \mathcal{K} \rangle$, where $\mathcal{N} = \{1, \cdots, N\}$ is the set of wireless users, which are the rational decision-makers in the system; $\mathcal{W}$ is the available network resources, e.g. frequency bands, time slots, or routes; $\mathcal{A}$ is the joint action space $\mathcal{A} = \times_{n \in \mathcal{N}} \mathcal{A}_n$, with $\mathcal{A}_n$ being the action set available for user $n$ to play the resource sharing game (e.g. selecting a specific frequency channel, time slot, or route for transmission); $\mathcal{U}$ is the utility vector function defined as a mapping from joint actions and the network resources to an $N$-dimensional real vector with each element being the utility of a particular user, i.e. $\mathcal{U} = \times_{n \in \mathcal{N}} u_n : \mathcal{A} \times \mathcal{W} \mapsto \mathbb{R}_+^N$; $\mathcal{K}$ represents the joint knowledge set of all the users. Specifically, for any $\boldsymbol{k} = (k_1, \cdots, k_N) \in \mathcal{K}$, $k_n$ indicates user $n$'s available knowledge about its opponents and the network resource. In a communication environment, users interact with each other and determine their actions such that they maximize their utility functions based on their available knowledge. Note that, as opposed to traditional strategic game definitions [26], we purposely include two new components $\mathcal{W}$ and $\mathcal{K}$ in the game formulation, because both of them play very important roles in communication games. Specifically, the available network resources $\mathcal{W}$ over which the game is played directly influences the utility achievable by the participating users. The next section will investigate the impact of knowledge $\mathcal{K}$ in communication games. The proposed game formulation is general and it can be applied in numerous communication scenarios. As we will see in the next section, many well-known game models are special cases of the above formulation.

## III. KNOWLEDGE IN COMMUNICATION GAMES

### A. *The Value of Knowledge*

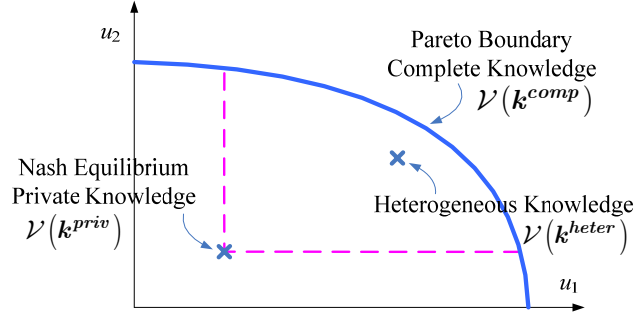

Fig. 3. The value of knowledge.

We define the policy with which user $n$ engages in the multi-user game $\pi_n : \mathcal{K}_n \mapsto \mathcal{A}_n$, as a mapping from the user's available knowledge into its specific selected action. The value of knowledge is defined as

$$\mathcal{V}(\boldsymbol{k}) = \left\{ \mathcal{U}\left( \pi_1(k_1), \cdots, \pi_N(k_N), w \right) \right\}, \tag{1}$$

which is the utility of individual users given their selected actions based on their knowledge[3]. This metric represents the resulting performance given the decentralized knowledge available to the users. However, as we will discuss later, existing research on communication games usually implicitly assumes that the knowledge vector $\boldsymbol{k}$ takes two extremes: either complete knowledge about the entire multi-user system $\boldsymbol{k}^{comp}$ or private knowledge $\boldsymbol{k}^{priv}$ about themselves. By explicitly allowing the knowledge $k_n$ in equation (1) to vary, i.e. individual users have different degrees of knowledge, we aim at providing a systematic characterization of the impact of knowledge availability in communication networks. Figure 3 shows an example of the value of heterogeneous and decentralized knowledge $\boldsymbol{k}^{heter}$ by comparing it with the conventional solution concepts, including Nash equilibrium and Pareto optimum that will be defined in the next subsection, where users have private knowledge $\boldsymbol{k}^{priv}$ or complete knowledge $\boldsymbol{k}^{comp}$, respectively.

---

[3] For the remaining part of this report, we omit $w$ in the expression of utility functions for static network resource.

Intuitively, users can benefit themselves by gaining more knowledge. However, this may potentially harm other users [49]. As we will show later, it is surprising that there are also settings where it can lead to $\mathcal{V}\left(\boldsymbol{k}^{heter}\right) \succ \mathcal{V}\left(\boldsymbol{k}^{priv}\right)^4$, if $\boldsymbol{k}^{heter} \supseteq \boldsymbol{k}^{priv}$, i.e. a subset of self-interested users with more knowledge may benefit the utilities of all users in the communication system.

### B. Special Cases: Private Knowledge and Complete Knowledge

Roughly speaking, the existing multi-user research can be categorized into two types, *non-cooperative* games and *cooperative* games. In wireless networks, any action taken by a single user usually affects the utilities of the other users sharing the same resources. Various game theoretic solutions were developed to characterize the resulting performance in both game models, among which the most well-known ones include *Nash equilibrium* (NE) and *Pareto optimality* [26]. As we will see, both scenarios implicitly assume that the users' available knowledge takes a homogeneous form. This is mainly because most research assumes that only users with homogeneous protocols, actions, etc. can interact. However, new systems are currently emerging, where heterogeneous users with different protocols and running different applications co-exist and influence each other.

First, non-cooperative approaches generally assume that the participating users only have private knowledge $k_n^{priv}$ about their own channel conditions, utility functions, etc. These users simply choose actions $a_n \in \mathcal{A}_n$ to selfishly maximize their individual utility functions. In other words, the policy of each user is $\pi_n(k_n^{priv}) = \arg\max\limits_{a_n \in \mathcal{A}_n} u_n\left(a_n, a_{-n}\right)$ in which $a_{-n} = (a_1, \cdots, a_{n-1}, a_{n+1}, \cdots, a_N)$. Most non-cooperative approaches are devoted to investigating the existence and properties of NE. NE is defined to be a profile $\left(a_1^*, \cdots, a_N^*\right)$ of actions with the property that for every player, it satisfies $u_n\left(a_n^*, a_{-n}^*\right) \geq u_n\left(a_n, a_{-n}^*\right)$ for all $a_n \in \mathcal{A}_n$, i.e. given the other users' actions, no user can increase its utility alone by changing its action. As shown in Figure 3, it is well-known that operating in non-cooperative manners will generally limit the performance of the device itself as well as the whole system, because the available resources are not always effectively exploited due to the conflicts of interest occurring among users [27].

---

$^4$ The symbol $\succ$ denotes component-wise inequality.

On the other hand, cooperative approaches focus on studying how users can jointly optimize a common objective function $f(u_1, \cdots, u_N)$. This function represents the social welfare (allocation rule) based on which the system-wide resource allocation is performed. Allocation rules, e.g. weighted sum maximization, can provide reasonable allocation outcomes by considering the trade-off between fairness and efficiency. However, this is true only when users agree on the allocation rules, or they are price-takers, which disregards the decentralized knowledge of users, and their strategic nature.

A profile of actions is Pareto optimal if there is no other profile of actions that makes every player at least as well off and at least one player strictly better off. Most cooperative approaches focus on studying how to efficiently find the optimum joint policy $a^* = \arg\max_{a \in \mathcal{A}} f(u_1, \cdots, u_N)$. It should be pointed out that, in order to achieve Pareto optimality, information exchange throughout the whole system is required such that all the users can collaboratively maximize $f(u_1, \cdots, u_N)$ and improve the system efficiency. These cooperative scenarios either assume complete knowledge $k^{comp}$ about all the users by a trusted moderator or peer (e.g. access point, base station, selected network leader etc.), to which it is given the authority to centrally divide the available resources among the participating users [11], or, in the distributed setting [7][12][13], users exchange price signals $\boldsymbol{p} = \{p_1, \cdots, p_N\}$ that reflect the "cost" for consuming the constrained resources and maintain the required knowledge of $k_n = \{u_n, p_n\}$ to maximize the social welfare and reach Pareto optimal allocations. In both centralized and decentralized scenarios, the users are assumed to have as objectives the maximization of social welfare. However, in many practical settings, strategic users aim at optimizing their own utilities based on their knowledge, rather than that of the system.

*C. An Example: Power Control Games with Decentralized and Heterogeneous Knowledge*

As discussed previously, existing approaches only assume two extremes for the users' knowledge. However, devices in current wireless networks can differ a lot in their abilities to sense the environment, to gather information, to reason about this information and proactively make decisions, which motivates us to investigate the impact of heterogeneity (in terms of action sets, utilities, knowledge) in the communication system. The decentralized heterogeneous knowledge of devices can take various forms, e.g. knowledge about resource allocation rules, available resource, opponents' actions, their strategies, and their utility functions. As an illustration, we investigate a multi-user power control game that considers how users should allocate their power in multi-carrier systems [10].

*1) Power Control Games*

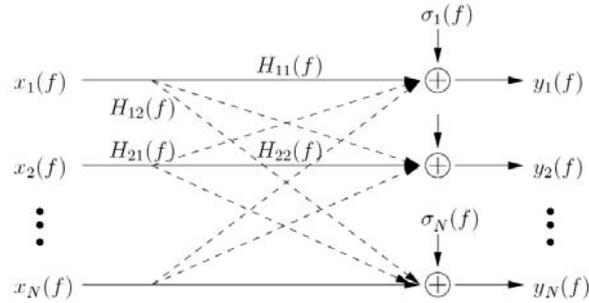

Fig. 4. Gaussian interference channel model.

Figure 4 shows the frequency-selective Gaussian interference channel models. There are $N$ transmitters and $N$ receivers in the system. Each transmitter and receiver pair can be viewed as a player. The transfer function of the channel from transmitter $i$ to receiver $j$ is denoted as $H_{ij}(f)$, where $0 \leq f \leq F_s$. The noise power spectral density (PSD) that receiver $n$ experiences is denoted as $\sigma_n(f)$ and its transmit PSD as $P_n(f)$. Its transmit PSD is subject to its power constraint:

$$\int_0^{F_s} P_n(f) df \leq \mathbf{P_n}. \tag{2}$$

For a fixed $P_n(f)$, if treating interference as noise, user $n$ can achieve the following data rate:

$$R_n = \int_0^{F_s} \log_2 \left( 1 + \frac{P_n(f)|H_{nn}(f)|^2}{\sigma_n(f) + \sum_{j \neq n} P_j(f)|H_{jn}(f)|^2} \right) df \, . \tag{3}$$

In the power control games, the set of players is $\mathcal{N} = \{1, \cdots, N\}$. The set $\mathcal{A}_n$ of actions available to user $n$ are the set of transmit PSDs satisfying the constraint in (2), and $u_n$ is user $n$'s achievable rate $R_n$. To fully capture the performance tradeoff in the system, the rate region is defined as

$$\mathcal{R} = \left\{ (R_1, \cdots, R_N) : \exists (P_1(f), \cdots, P_N(f)) \; satisfying \; (2) \; and \; (3) \right\}, \tag{4}$$

which contains all the achievable data rate combinations among the users subject to the power constraints.

In the following subsections, we will investigate the power control games with homogeneous and heterogeneous users, and show that self-interested users can potentially improve the system performance if they have more knowledge.

### 2) Power Control Game with Homogeneous Users

Both non-cooperative and cooperative solutions have been developed for power control games. First, a non-cooperative iterative water-filling (IW) algorithm has been proposed to optimize the performance without the need for a central controller [10]. It assumes that each user only has the knowledge of its channel gain and the noise-and-interference PSD, i.e. $k_n^{priv} = \left\{ |H_{nn}(f)|^2, \right.$ $\left. \sigma_n(f) + \sum_{j \neq n} P_j(f)|H_{jn}(f)|^2 \right\}$. At every decision stage, self-interested users deploying this algorithm try to myopically maximize their immediate achievable rates. Specifically, for a two-user system, each user simply updates their transmit PSD by choosing the optimal solutions of

$$\max_{P_1(f)} \int_0^{F_s} \ln \left( 1 + \frac{|H_{11}(f)|^2 P_1(f)}{\sigma_1(f) + |H_{21}(f)|^2 P_2(f)} \right) df$$
$$s.t. \; \int_0^{F_s} P_1(f) \, df \leq \mathsf{P}_1 \tag{5}$$
$$P_1(f) \geq 0, \; \; \forall f$$

and

$$\max_{P_2(f)} \int_0^{F_s} \ln\left(1 + \frac{\left|H_{22}(f)\right|^2 P_2(f)}{\sigma_2(f) + \left|H_{12}(f)\right|^2 P_1(f)}\right) df$$
$$s.t. \int_0^{F_s} P_2(f)\, df \le \mathsf{P_2} \qquad\qquad . \tag{6}$$
$$P_2(f) \ge 0, \ \ \forall f$$

Both users will iteratively perform water-filling with respect to their experienced noise-and-interference PSD $\sigma_n(f) + \sum_{j\ne n} P_j(f)\left|H_{jn}(f)\right|^2$ across the whole frequency band until a Nash equilibrium is reached.

On the other hand, cooperative solutions in power control games can attain Pareto optimality by sharing their complete knowledge $k^{comp} = \left\{\left\{H_{ij}(f)\right\}, \left\{\sigma_n(f)\right\}\right\}$ throughout the system and solving the optimization globally [11]. To find the Pareto optimal operating points on the boundary of the convex hull of the rate region $\mathcal{R}$, the cooperative approaches focus on solving the weighted rate-sum maximization

$$\max_{P_1(f),\cdots,P_N(f)} \sum_{n=1}^{N} w_n \int_0^{F_s} \log_2\left(1 + \frac{P_n(f)\left|H_{nn}(f)\right|^2}{\sigma_n(f) + \sum_{j\ne n} P_j(f)\left|H_{jn}(f)\right|^2}\right) df$$
$$s.t. \int_0^{F_s} P_n(f)\, df \le \mathsf{P_n} \ \ \forall n \qquad\qquad , \tag{7}$$
$$P_n(f) \ge 0, \qquad\qquad \forall n, f$$

in which the $n$ th element in the weighted vector $\boldsymbol{w} = \left[w_1 \cdots w_N\right]$ indicates the relative importance of the user $n$ 's utility [11][12].

Figure 5 shows the achievable rate region for both approaches [12]. Similar with Figure 3, we can see that there is substantial performance gain obtained by the cooperative algorithms over the non-cooperative IW algorithm. However, all these algorithms still assume homogeneous participants in the sense that they deploy the same algorithms and have the similar information availability. Therefore, they are not well-suited to be applied in the scenarios where users differ in their ability to acquire information and improve performance.

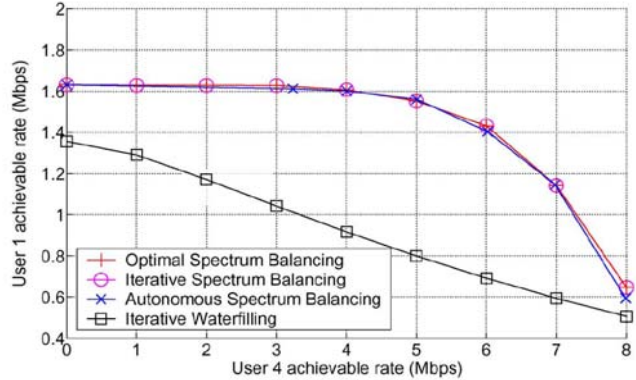

Fig. 5. Rate regions obtained by various algorithms [12].

### 3) *Heterogeneous Knowledge and Stackelberg Equilibrium*

To give an example of how users with heterogeneous knowledge can improve the system performance, we introduce a simple two-user two-channel power control game, where $H_{11} = H_{22} = 1$ and $\sigma_1 = \sigma_2 = 1$ for both channels, $H_{12}(1) = 0.8$ and $H_{12}(2) = H_{21}(1) = H_{21}(2) = 0.4$. Users' power constraints are $\mathsf{P_1} = \mathsf{P_2} = 10$. User 1 has two actions including *Spread* and *Concentrate*. In the *Spread* mode, user 1 will split its power in both channels. In the *Concentrate* mode, it will concentrate all its power in channel 1. User 2 has two similar actions, whereas it will transmit at its maximum power instead in channel 2 in the *Concentrate* mode. If users choose a (*Concentrate*, *Spread*) play, user 1's utility will be $\log_2\left(1 + \dfrac{10}{1 + 0.4*5}\right) = 2.12$ and user 2's is

$\log_2\left(1 + \dfrac{5}{1 + 0.8*10}\right) + \log_2(1 + 5) = 3.22$. Similarly, we can calculate users' utilities for all the action combinations and the payoff matrix of this game is shown in Figure 6. Note that in this game, user 1 has a strictly dominant strategy, *Spread*. Therefore, two players will end up with a (*Spread*, *Spread*) play if they have only the private knowledge and will always take the best response. The resulting outcome in this case will be $(2.83, 2.42)$. If user 1 is aware of user 2's coupled reaction, it will play *Concentrate* instead, such that they will end up with a (*Concentrate*, *Concentrate*) play, which leads to an increased utility of $(3.46, 3.46)$ for both players. It is worth noticing that additional knowledge is needed to attain this performance improvement, i.e. user 1 needs to know the utility and the response strategy of the other user.

| User 1 \ User 2 | *Spread* | *Concentrate* |
|---|---|---|
| *Concentrate* | 2.12, 3.22 | 3.46, 3.46 |
| *Spread* | 2.83, 2.42 | 3.59, 2.12 |

Fig. 6. A simple power control game: user 1's utility is given first in each cell, with user 2's following.

To address heterogeneity in the above case, we introduce the concept of Stackelberg equilibrium [2].

**Definition of Stackelberg Equilibrium**: Let $NE(a_n)$ be the Nash equilibrium strategy of the remaining players if player $n$ chooses to play $a_n$. The strategy profile $\left(a_n^*, NE\left(a_n^*\right)\right)$ is a Stackelberg equilibrium for user $n$ iff

$$u_n\left(a_n^*, NE\left(a_n^*\right)\right) \geq u_n\left(a_n, NE\left(a_n\right)\right), \forall a_n \in \mathcal{A}_n.$$

### 4) Power Control Game with Heterogeneous Users

In contrast with existing approaches, we study a power control game with two heterogeneous users [28]. Specifically, user 2 has only the knowledge $k_2^{priv} = \left\{\left|H_{22}(f)\right|^2, \sigma_2(f) + P_1(f)\left|H_{12}(f)\right|^2\right\}$ and it always takes the best response strategy of water-filling. The knowledge of user 1 $k_1^{heter}$ includes $\left\{H_{ij}(f)\right\}, \left\{\sigma_k(f)\right\}$, and user 2's policy $\pi_2\left(k_2^{priv}\right)$ of water-filling. We apply the concept of Stackelberg equilibrium to determine user 1's optimal action by formulating the following bi-level programming problem:

$$\begin{cases} \textit{upper level problem} \begin{cases} \max\limits_{P_1(f)} \int_0^{F_s} \ln\left(1 + \dfrac{\left|H_{11}(f)\right|^2 P_1(f)}{\sigma_1(f) + \left|H_{21}(f)\right|^2 P_2(f)}\right) df \\ s.t. \ \int_0^{F_s} P_1(f)\, df \leq \mathsf{P}_1 \\ P_1(f) \geq 0 \end{cases} \\ \textit{lower level problem} \begin{cases} P_2(f) = \arg\max\limits_{P_2'(f)} \int_0^{F_s} \ln\left(1 + \dfrac{\left|H_{22}(f)\right|^2 P_2'(f)}{\sigma_2(f) + \left|H_{21}(f)\right|^2 P_1(f)}\right) df \\ s.t. \ P_2'(f) \geq 0 \\ \int_0^{F_s} P_2'(f)\, df \leq \mathsf{P}_2 \end{cases} \end{cases} \quad (8)$$

As indicated in (8), user 1 should optimize its transmit PSD by exploring its knowledge and including the opponent's policy $\pi_2\left(k_2^{priv}\right)$ as a constraint in the lower level problem in the bi-level program, whereas in the iterative water-filling algorithm, user 1 will solve the upper level problem instead since it only has the private knowledge.

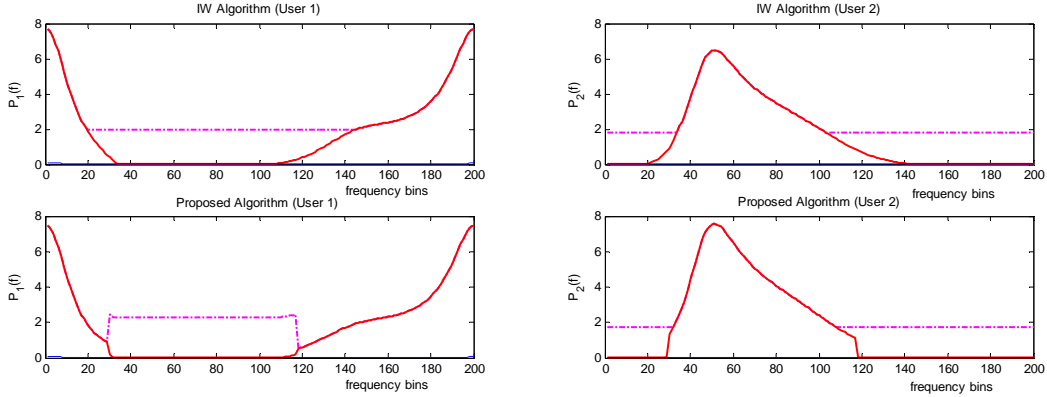

Fig. 7. Power allocations of different algorithms.

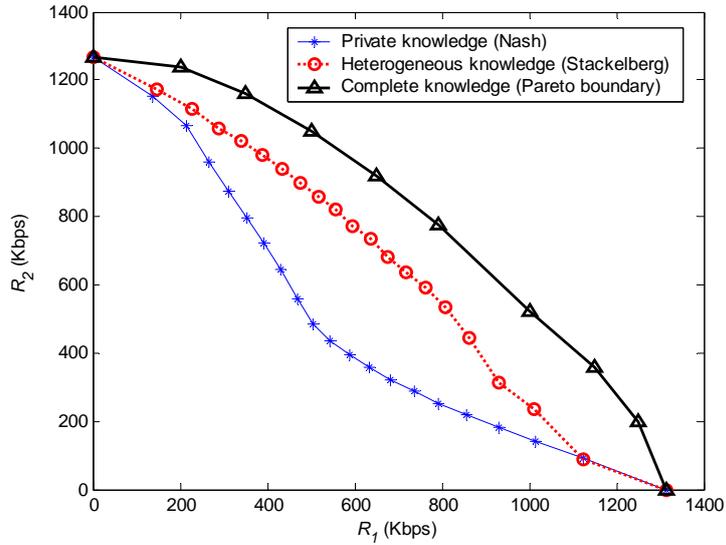

Fig. 8. Value of heterogeneous knowledge in power control game.

We provide a sub-optimal solution of problem (8) to verify the performance improvement when having heterogeneous knowledge. Figure 7 shows the power allocations for both users using iterative water-filling algorithm and our proposed algorithm. In the iterative water-filling algorithm, each user water-fills the whole frequency band by regarding its competitor's interference as background noise. In contrast, user 1 will not water-fill, if it has additional knowledge $k_1^{heter}$ of user 2's channel state information and power allocation strategy. As shown in Figure 7, user 1 concentrates its power in the interval $[30,117]$ even though it can gain an immediate increase in $R_1$ by re-allocating some of its power in the region where the noise PSD is below its water-level, e.g. $[20,30]$ and $[117,140]$.

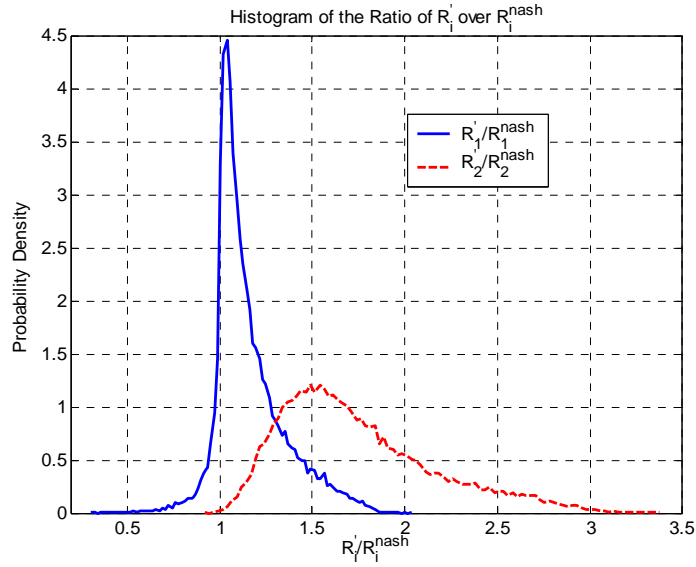

Fig. 9. Histogram for the ratio of $R_1' / R_1^{nash}$ and $R_2' / R_2^{nash}$

To evaluate the performance, we tested a variety of frequency-selective fading channels where the iterative water-filling algorithm reaches the Nash equilibrium. The total power of all rays of $H_{11}(f)$ and $H_{22}(f)$ is normalized as one, and that of $H_{12}(f)$ and $H_{21}(f)$ is normalized as $0.5$. Figure 8 shows the achievable rate regions of different algorithms for one realization of randomly generated channels, which represent the values of various knowledge, including $k^{priv}$, $k^{heter}$, and $k^{comp}$, in the two-user power control games with different combinations of power constraints. We can see the achievable rate region is enlarged in the case with heterogeneous users compared with the iterative water-filling solution. In other words, as illustrated in Figure 3, user 1's additional knowledge can benefit both of them even though user 2 behaves in a non-collaborative manner. Figure 9 shows the simulated histogram of the ratio of $R_1' / R_1^{nash}$ and $R_2' / R_2^{nash}$, where $R_i'$ is user $i$'s achievable rate of our proposed algorithm in the heterogeneous case and $R_i^{nash}$ represents that of user $i$ if both of them adopt the water-filling algorithm. Simulation results show that both users will gain rate improvement in most of the realizations. The average rate improvement for user 1 is 16.4%. Interestingly, the average rate improvement for user 2 is 74%, which is significantly higher than that of user 1.

## IV. STRATEGIC LEARNING IN COMMUNICATION GAMES

In the previous section, we define the value of knowledge, which represents the achievable performance bound when heterogeneous users with different amounts of knowledge interact and reach steady-state. In this section, we will discuss how users can approach these bounds using strategic learning.

### A. Learning in Communication Games

Most non-cooperative game theory research focuses on studying various equilibrium concepts. The traditional explanation for when and why equilibrium arises is that it results from analysis and introspection by the players in a situation where the rules of the game, the rationality of the players, and the players' payoff functions are all common knowledge. However, this is difficult to attain in communication networks, where the information is decentralized and often private to the users. Thus, we use another strand of game theory known as "learning in games" [29] in order to construct operational solutions that allow autonomous users to approach the performance bounds of knowledge that they try to obtain via strategic learning. In contrast with issues such as equilibrium characterization, the area of learning in games focuses on developing constructive methods to achieve various points in the utility space, where agents may or may not be person-by-person optimal. It also provides an alternative explanation of how equilibrium arises, as the long-run outcome of a process in which less than fully rational players grope for optimality over time. The main interest of strategic learning is to understand what will be the emergent behavior of interacting agents that use simple adaptation rules to adjust and optimize their strategies. The outcome of these interactions need not converge to equilibrium, i.e., perpetual adaptation of strategies may persist, but the user's and the system's utilities may substantially improve compared to the performance of NE. While originally learning in games was used to develop descriptive models in social systems [47], these methods can also be used as a prescriptive model driving the design of dynamic reconfiguration mechanisms in multi-agent systems. This has motivated recently a growing interest in the application of strategic learning to engineered systems (e.g. in robotics [30]-[32] etc.). However, strategic learning algorithms have only been recently used in communication environments [33]-[41].

In the previous section, we demonstrated the incentives of self-interested users to increase their knowledge. However, the process of acquiring the desired knowledge can be very difficult in general, because the intentions of users could be difficult to predict, and their knowledge and skills could be difficult to envisage. Thus, a strategic user can only predict these uncertainties caused by the competing users based on its explicit information exchanges with other users and its observations of past interaction. To address this challenge, a user can deploy strategic learning solutions to build beliefs based on their available information and, based on these beliefs, optimize their actions. Strategic learning algorithms have the following benefits when applied in multi-user communications settings. First, strategic learning can address the challenges arising due to the distributed availability of information and knowledge in communication networks, where the decisions on how to adapt the transmission strategies need to be performed in a distributed manner, as the delay caused by information exchanges and communication overheads usually do not allow propagating messages back and forth to a central decision maker. Second, as long as communication devices are built according to specific protocols, users' behaviors, actions and even payoff functions, can be parameterized. As a result, users are able to easily model their opponents, and strategically learn the desired knowledge, thereby improving their performance through repeated interactions. Third, if we consider dynamic environments, the operating conditions of interacting users are time-varying, and thus, they need to timely adapt to the network dynamics. In short, strategic learning is an appealing approach for driving the information acquisition and decision making in multi-user communication systems, where users need to optimize their transmission strategies in an informationally decentralized manner.

The existing learning in games literature provides a broad spectrum of analytical and practical results on learning algorithms and underlying game structures for a variety of competitive interaction scenarios. In general, the main issue considered has been to characterize long term behavior in terms of a generalized equilibrium concept or characterize the *lack* of convergence for general classes of learning dynamics. However, when selecting learning solutions for communication games, the specific constraints and features of the considered multi-user interactions will need to be considered. For instance, the learning algorithm that should be deployed by a user in a wireless environment strongly depends on what information it can observe about the other users, given the adopted protocols or spectrum regulation rules. In this report, we assume that knowledge can be gathered by users based on observations and/or information. We differentiate these two types of knowledge gathering solutions, since we want to highlight the different communication overheads incurred as well as the challenges that arise when heterogeneous users, deploying different protocols and/or having different learning abilities interact.

There are several different forms of information in communication systems:

- *Private information*: this private information can include the utility forms, characteristics of the application traffic, channel gains or channel conditions (SINR, etc.);

- *Network information*: the network information refers to the network resource states (e.g. which channels are available for transmission);

- *Opponents' information*: this information can include the protocols, action sets, strategies, and even utilities of the opponents.

*B. Definition of Learning Algorithms and Beliefs in Communication Games*

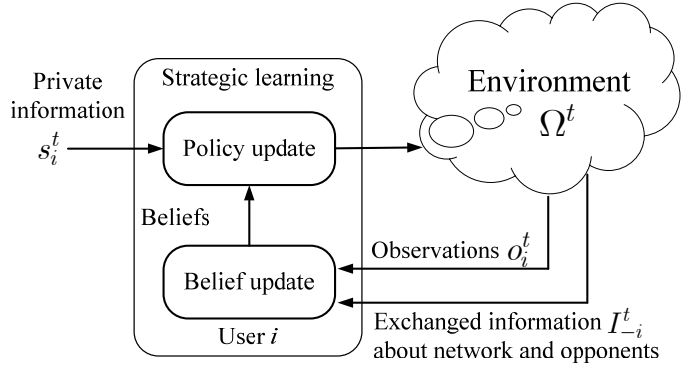

Fig. 10. An illustration of strategic learning

We note that a learning algorithm is built based on all the observation $o_i^t$ and exchanged information $I_{-i}^t$ obtained by user $i$, and hence, it is denoted as $\mathcal{L}_i(o_i, I_{-i})$. Based on its available information and observations, a user can build its *belief* about other users' strategies, policies, and the network resource, and update its response policy. As shown in Figure 10, in communication environment, a multi-agent strategic learning algorithm $\mathcal{L} = \times_{i \in \mathcal{N}} \mathcal{L}_i$ can be defined using the following equations:

$$a_i^t = \pi_i^t\left(\hat{k}_i^t\right), \, with \, \hat{k}_i^t = \left[ s_i^t \,\, B_{s_{-i}}^t \,\, B_{\pi_{-i}}^t \,\, B_w^t \right], \tag{9}$$

$$\Omega^t = Game\left(\mathcal{N}, w^t, \boldsymbol{a}^t, \boldsymbol{u}^t, \hat{\boldsymbol{k}}^t\right), \tag{10}$$

$$\boldsymbol{o}^t = \times_{i \in \mathcal{N}} o_i^t = O\left(\Omega^t\right), \tag{11}$$

$$\pi_i^{t+1} = \mathcal{F}_i\left(\pi_i^t, \hat{k}_i^t\right), \tag{12}$$

$$B_{\pi_{-i}}^{t+1} = \mathcal{F}_{\pi_{-i}}\left(B_{\pi_{-i}}^t, o_i^t, I_{-i}^t\right), \,\, B_w^{t+1} = \mathcal{F}_w\left(B_w^t, o_i^t, I_{-i}^t\right), \,\, B_{s_{-i}}^{t+1} = \mathcal{F}_{s_{-i}}\left(B_{s_{-i}}^t, o_i^t, I_{-i}^t\right), \tag{13}$$

where $\pi_i^t$ is the policy taken by user $i$ to select its action, $s_i^t$ represents the status of user $i$'s private information, $B_{s_{-i}}^t$, $B_{\pi_{-i}}^t$ and $B_w^t$ are the beliefs about other users' private information $s_{-i}$, policies $\boldsymbol{\pi}_{-i}$ and network resource $w$, $\hat{k}_i^t$ approximates the true knowledge $k_i^t$, and $I_{-i}^t$ is the explicit information that user $i$ has received about other users and the network resource; $\Omega^t$ is the output of the multi-user interaction; $o_i^t$ is the observation of user $i$ and $O$ is the observation function, which depends on the current game output; $\mathcal{F}_{\pi_{-i}}$, $\mathcal{F}_w$, and $\mathcal{F}_{s_{-i}}$ are the update functions about the policies and beliefs. Note that different learning algorithms can employ various update functions, and the three update functions listed above are not necessarily included.

Eq. (9) shows that user $i$ takes action based on its own private information, the belief about the other users' private information, policies and available network resources. After each user determines its actions, a multi-user communication game is played and the results of the game are produced as shown in Eq. (10). The results of the multi-user game may or may not be fully observed by the users. Eq. (11) represents the observation function which depends on the implemented network protocols and sensing and measuring abilities of users. Hence, a user may have incentives to exchange information with other users. Eq. (12) shows that a user updates its policy based on its private information and beliefs. The exchanged information $I^t_{-i}$ may be used to update the belief about the other users' states, policies and the network resource state. Eq. (13) represents the updates of these beliefs.

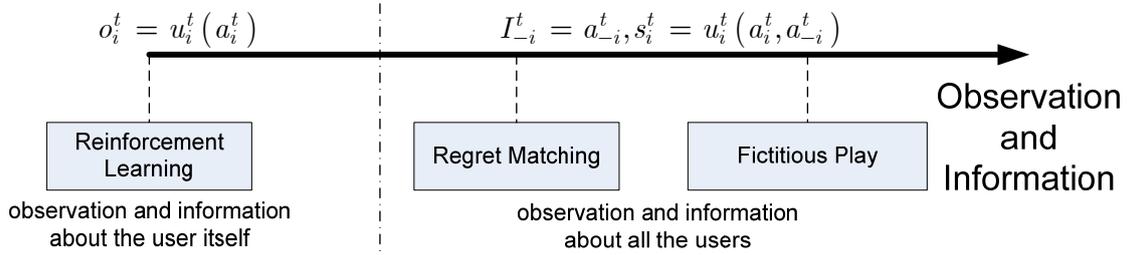

Fig. 11. Illustration of strategic learning solutions based on the required observation and information

An important consideration in strategic learning solutions is the observation $o^t_i$ and the information $I^t_{-i}$ available to each user as well as their ability to process this information. Figure 11 highlights three conventional strategic learning solutions [29][42]: 1) *Fictitious play*: a wireless user's strategy is the best response to the empirically observed actions of all other users ; 2) *Regret matching*: The user's strategies are selected to eliminate the "regret" of not having played different actions based on retrospective observations of other user's strategies; 3) *Reinforcement learning*: Playing a strategy increases or decreases the probability of it being played in the future, depending on the utility received when playing that action.

As shown in Figure 11, both regret matching and fictitious play require the information of the functional expression of utility $u_i^t \left( a_i^t, a_{-i}^t \right)$ and assume that the actions $a_{-i}^t$ of other users can be observed or explicitly exchanged. Of course, this may not be feasible, especially for "large scale" games that involve many users, or when the other users adopt different protocols. In contrast, reinforcement learning only assumes knowledge of the received payoff by individual users of their own historical actions $u_i^t \left( a_i^t \right)$, and it is easier to implement in wireless networks, where users with different protocols operate, and thus users cannot easily exchange information or interpret their opponents' strategies based on their observations. Another difference among these learning methods is their required complexity, which is especially important if deployed by low-power wireless devices. Hence, depending on their available observation and information, strategic users can deploy different learning solutions to achieve various performance versus complexity trade-offs.

If users deploy the same set of actions, strategies, and protocols, and are able to observe the other users' actions, they can model the other users using "self-play" [47] and hence, they can easily develop strategies for efficiently sharing the network resources by anticipating the other users' actions. However, if the heterogeneous users are not able to observe or interpret the other users' actions (e.g. because they deploy different protocols), they may share the available network resources in a very inefficient manner, as previously discussed. Hence, users may decide that it is beneficial for them to exchange explicit information about their actions, and potentially even their strategies or utilities [40]. (Such information exchanges can be implemented using existing or new signaling protocols [1] at the application or MAC layer.) Therefore, users need to decide which strategic learning solution, information and observations have higher values for them. Next, we will discuss how users can quantify the value of learning, information, and observations.

*C. Value of Learning, Information, and Observations*

As mentioned previously, the performance of a learning algorithm can be quantified based on the resulting payoffs obtained by users. We denote the policy generated by a learning algorithm $\mathcal{L}$ as $\boldsymbol{\pi}^{\mathcal{L}}$. Users will learn in order to improve their policies and payoffs from participating in the communication game. Given the available information and observations, the value of the learning algorithm $\mathcal{L}$ is defined as the time average payoff obtained in a time window with length $T$ in which this learning algorithm was used:

$$\mathcal{V}_{\mathcal{L}}(\boldsymbol{o}, \boldsymbol{I}, \boldsymbol{s}, T) = \frac{1}{T} \sum_{t=1}^{T} \mathcal{U}^t(\boldsymbol{\pi}^{\mathcal{L}(\boldsymbol{o}, \boldsymbol{I}, \boldsymbol{s})}), \qquad (14)$$

where the attainable payoff of the learning approach $\mathcal{L}$ depends on the observation $\boldsymbol{o} = \times_{i \in \mathcal{N}} o_i$, private information $\boldsymbol{s} = \times_{i \in \mathcal{N}} s_i$, and exchanged information $\boldsymbol{I} = \times_{i \in \mathcal{N}} I_{-i}$. Thus, using this definition, the value of a learning scheme can be determined. As illustrated in Figure 12, different learning algorithms provide operational ways to approach the value of perfectly having the knowledge that the users are trying to learn. The "value of information/observations" for various available information and observations with respect to a learning algorithm $\mathcal{L}$ can be also similarly computed, which will play a significant role on determining what information/observations should be exchanged/gathered by users.

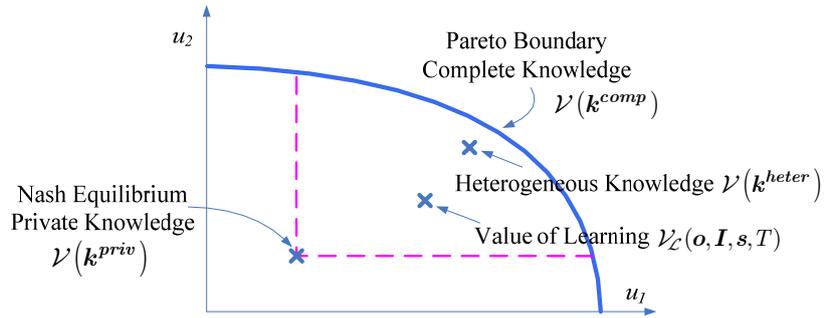

Fig. 12. An illustration of the value of knowledge and learning.

## D. An Example: Regret Matching

In this subsection, we introduce the concept of correlated equilibrium, which can be achieved using regret matching [42]. As an illustration, we also numerically show that users that employ this strategic learning algorithm can improve their performance without any explicit information exchange with their competitors in the power control games.

*1) Correlated Equilibrium*

| User 1 ＼ User 2 | *Aggress* | *Backoff* |
|---|---|---|
| *Aggress* | 0, 0 | 7, 2 |
| *Backoff* | 2, 7 | 6, 6 |

Fig. 13. Game in contention networks: user 1's utility is given first in each cell, with user 2's following.

First, we consider the game in contention networks, e.g. networks deploying CSMA/CA based protocols, where two users are competing with each other in a packet transmission contest where each of them can either aggressively transmit or obey the backoff principle [17]. If one user is going to *Aggress*, it is better for the other to *Backoff*, otherwise the network will be congested and few packets can get through. But if one user chooses to *Backoff*, it is better for the other to *Aggress*, which will result in a better outcome than both users obeying the backoff principle. The payoff matrix is shown in Figure 13. There are three NE in this game, including two pure strategy NE, (*Aggress*, *Backoff*) and (*Backoff*, *Aggress*), and a mixed strategy NE where both *Aggress* with probability 1/3 and get an average utility of 4.67.

Next, we introduce the concept of *correlated equilibrium*, which is a generalization of the NE concept. The correlated equilibrium is defined in a context where the players are able to access certain common signals. These signals allow players to coordinate their actions and to perform joint randomization over strategies. Given the recommended strategy, it is in the players' best interests to conform with this strategy. The distribution of the strategies is called correlated equilibrium.

***Definition of Correlated Equilibrium***: Let $\Delta(\mathcal{A})$ be the set of probability distributions on $\mathcal{A}$. A probability distribution on the strategy profile $\mu \in \Delta(\mathcal{A})$ is a correlated equilibrium (CE) of game $\Omega$ if and only if, for all $n \in \mathcal{N}$ and $(a_n, a_{-n}) \in \mathcal{A}$,

$$\sum_{a_{-n} \in \mathcal{A}_{-n}} \mu(a_n, a_{-n}) u_n(a_n, a_{-n}) \geq \sum_{a_{-n} \in \mathcal{A}_{-n}} \mu(a_n, a_{-n}) u_n(a'_n, a_{-n}) \text{ , for all } a'_n \in \mathcal{A}_n$$

A simple example of correlated equilibrium for the contention game is to let the two users choose

(*Backoff*, *Backoff*), (*Aggress*, *Backoff*), and (*Backoff*, *Aggress*) with identical probability of 1/3. Note the resulting utility for each user is 5, which is higher than that of the mixed strategy NE. For finite games, the set of CE is a non-empty polytope which contains the convex hull of all the NE. Figure 14 shows geometry of the equilibria of the 2-user contention game, where the tetrahedron is the simplex of probability distributions on outcomes of the game, the saddle is the set of distributions independent between players, the polytope with 5 vertices and 6 facets is the set of CE, and their three points of intersection are NE [43]. Importantly, as shown in Figure 14, since CE is a generalization of NE, a CE may lie outside the convex hull of the NE. Therefore, the performance of CE may be better than that of NE.

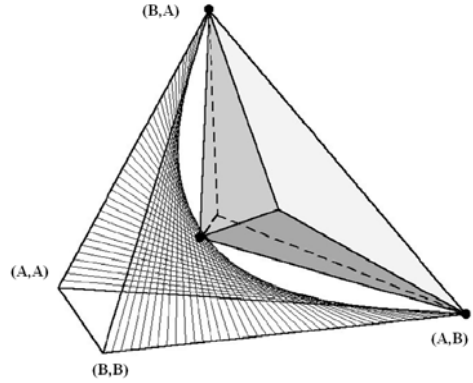

Fig. 14. Geometry of CE and NE (A=*Aggress*, B=*Backoff*) [43].

### 2) *Regret matching in Power Control Games*

To achieve the set of CE, a class of algorithms named regret matching was designed. The stationary solution of this algorithm exhibits no regret and users update their play probabilities in (9)-(13) are proportional to the "regrets" for not having played other actions [42]. In particular, for any two distinct actions $a'_n \neq a^t_n$ in $\mathcal{A}_n$ and at every time $t$, the regret of user $n$ at time $t$ for not playing $a'_n$ is

$$r^t_n\left(a'_n\right) = \max\left\{0, \frac{1}{t}\sum_{t' \leq t} u_n\left(a'_n, a^{t'}_{-n}\right) - u_n\left(a^{t'}_n, a^{t'}_{-n}\right)\right\}. \tag{15}$$

In other words, the regret for an action is defined as the increase in utility if such a change had always been made in the past. Computing the regret completes the belief update part in the learning process in Figure 10. Users will update their policies by simply assigning the probability of selecting a certain transmission action proportionally to the regret of that action. It is shown that, if every player plays according to regret matching algorithm, the empirical distributions of play converge almost surely to the set of CE distributions of the game [42]. Specifically, in regret matching, the observation $o_i^t$ in the past play serves the role of common signal that enables users to perform joint coordination. The advantage of the algorithm is that it requires no information exchange $I_{-i}^t$ among the users, and hence users can learn the correlated equilibrium in a fully distributed manner.

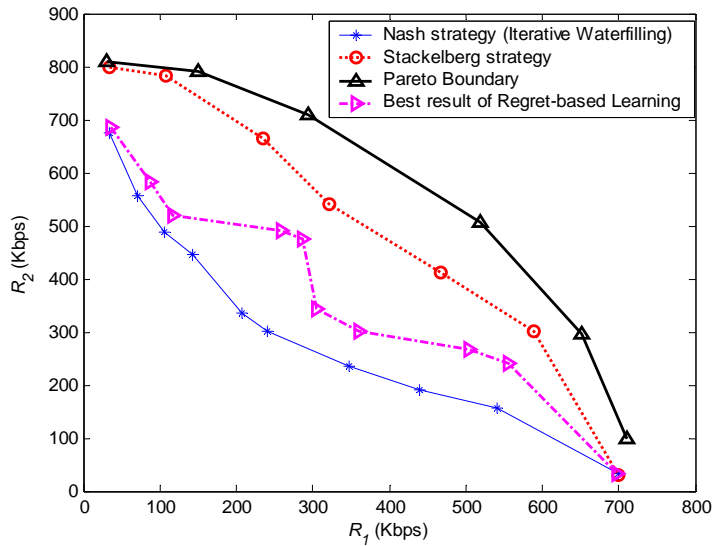

Fig. 15. Performance of regret matching in power control games.

We applied this regret matching algorithm in a two-user power control game and the simulation results are shown in Figure 15. In the simulation, we discretize users' action space by dividing their maximum power into multiple parts. Each user calculates the regrets for all the possible actions in its finite action space and updates its strategy repeatedly according to (15). The numerical results for different power constraints are shown in Figure 15. Because NE is only part of the set of CE, the best performance that regret matching can achieve is strictly better than that of NE achieved by the

IW algorithm.

## V. Some Challenges for Future Research

Numerous challenges still need to be addressed to characterize the performance of the knowledge-driven games among heterogeneous users as well as to make these solutions widely adopted in practice. For example, the value of having different knowledge (e.g. the knowledge of other users' action space, strategies, utility functions, the environment, etc.) needs to be quantified in various communication systems, and application-specific learning algorithms (e.g. which consider the available common knowledge about the protocols present in the system) need to be designed to achieve an increased utility efficiency instead of applying learning algorithms designed for generic games. Moreover, in each application scenario, the learning algorithms need to be benchmarked along the following three dimensions: (i) incurred communication overhead for the information exchanges, (ii) incurred computational overhead, and (iii) the achievable utility that can be derived by different learning algorithms (i.e. the value of learning).

Another interesting topic is to investigate the scenarios in which users only have the (probabilistic) belief about the other players' characteristics. In these cases, Bayesian games are an appealing approach to model the interaction and determine the achievable performance [26].

Considering the dynamic nature of the wireless networks where users' CSI and source characteristics are dynamically changing over time, all the previously addressed issues will be further complicated. More advanced multi-user interaction models, e.g. stochastic games, are required to cope with these dynamics in the multi-user interaction [41].

Research advances will also need to be made to allow users to proactively improve the available common knowledge based on which the multi-user interact, and design protocols that can encourage the accumulation of common knowledge, which can increase the system performance and even prevent malicious users to misbehave.

Summarizing, in this report, we have provided a fresh glimpse of how heterogeneous knowledge and strategic learning can change the conventional way people have designed and characterized the multi-user interactions in wireless communication systems. Considering the complexity of this research topic and the diversity of existing literature, our presentation is far from comprehensive. However, we hope that this report can motivate the need for further research activities in studying the knowledge-driven interaction of users in communication networks. Finally, it is our vision that designing communication systems, where users with different amounts of knowledge can compete according to "free market" principles rather than based on guidelines imposed by resource owners and protocol providers, will lead to unprecedented performance improvements for communication devices and systems as well as catalyze new algorithm and system designs.